\documentclass[12pt]{article}
\usepackage{epsfig}
\usepackage{amssymb}
\usepackage{graphics}
\usepackage{amsmath,amsfonts,amssymb,graphicx}
\usepackage{flushend,cuted}
\newcommand {\be}{\begin{equation}}
\newcommand {\ee}{\end{equation}}
\newcommand{\beq}{\begin{eqnarray}}
\newcommand{\eeq}{\end{eqnarray}}

\begin{document}
\pagestyle{plain} \baselineskip 0.4cm \vspace{0.1cm}
\begin{center}
\Large \noindent {\bf Temperature dependence of Quark and Gluon
condensate In The Dyson-Schwinger Equations At Finite Temperature}

\footnote[1]{{ The work was supported in part by National Natural
Science Foundation of China (11365002), Guangxi Natural Science
Foundation for Young Researchers (2013GXNSFBB053007), Guangxi
Education Department (2013ZD049), Guangxi Grand for Excellent
Researchers (2011-54), the Funds of Guangxi University of Science
and Technology for Doctors (11Z16).}}
\end{center}
\vspace{1.2 cm}
\begin{center}
{\bf Zhou Li-Juan$^{1}$, Zheng Bo$^{1}$, Zhong Hong-wei$^{1}$,  Ma Wei-xing$^{2}$}  \\

\vspace{0.4cm} $^{1}$ School of Science, Guangxi University of
Science and Technology, Liu Zhou, 545006, China
\\
\vspace{0.4cm}
$^{2}$ Institute of High Energy Physics, Chinese
Academy of Sciences, Beijing, 100049, China
\end{center}
\vspace{0.4cm}
\begin{center}
{\Large \bf Abstract}
\end{center}
Based on the Dyson-Schwinger Equations (DSEs) with zero- and finite temperature, the two quark
condensate, the four quark condensate and quark gluon mixed condensate in non-perturbative QCD
state are investigated by solving the DSEs respectively at zero and finite temperature. These
condensates are important input parameters in QCD sum rule with zero and finite temperature
and properties of hadronic study. The calculated results manifest that the three condensates
 are almost independent of the temperature below the critical point temperature $T_{c}$.
 The results also show that the chiral symmetry restoration is obtained above $T_{c}$.
 At the same time, we also calculate the ratio of the quark gluon mixed condensate to
 the two quark condensate which could be quark virtuality. The calculations show that
 the ratio $m^{2}_{0}(T)$ is almost flat in the region of temperature from $0$ to $T_{c}$,
 although there are drastic changes of the quark condensate and the quark gluon mixed
 condensate at this region of $T_{c}$.The predicted ratio comes out to be $m^{2}_{0}(T)= 2.41GeV^{2}$
 for vacuum state at the Chiral limit, which
suggests the significance that the quark gluon mixed condensate has
played in OPE.

\vspace{1.0cm} \noindent {\bf Key words:} Dyson-Schwinger Equations at zero and finite temperature,
Dynamical chiral symmetry breaking, Quark and gluon condensate.\\

{\bf PACS Number(s): 12.38.Lg, 12.38.Mh, 24.85.+p}

\section{Introduction}

With the development of heavy-ion collision experiments, more
attentions have been turning to exploring the hot and dense QCD
matter. The hot and dense matter can be studied via various
approaches, such as: lattice QCD, QCD sum rules, Chiral perturbation
theory as well as the Dyson-Schwinger equations (DSEs) and so on.
Due to the asymptotic freedom feature of QCD, the QCD matter will
take place a phase transition from hadronic phase, with quarks and
gluons being bound states inside hadron, to the quark gluon plasma
phase where the bound clusters of quarks and gluons have been
de-confined at sufficient high temperature and /or density. Studying
the Chiral condensates at zero- and finite temperature is a crucial
importance of nuclear and hadronic physics research, even if for
astrophsics and cosmology study.

According to QCD sum rules, quarks exist in the vacuum of
non-perturbative QCD, which is densely populated by long-wave
fluctuations of gluon fields. The order parameters of this
complicated state are described by various vacuum condensates
$\langle 0 \mid : \bar{q}q : \mid 0 \rangle$ , $\langle 0 \mid
:G^{a}_{\mu\nu}G^{a}_{\mu\nu} : \mid 0 \rangle$, $\langle 0
\mid:\bar{q}[ig_{s}G^{a}_{\mu\nu}\sigma_{\mu\nu}
\frac{\lambda^{a}}{2}] q : \mid 0 \rangle$, \ldots, which are the
vacuum matrix elements of various singlet combinations of quark and
gluon fields. In QCD sum rules, various condensates are input
parameters so that they play an important role to reproduce various
hadronic properties phenomenologically in the operator product
expansion calculations
(OPE)\cite{Shifman:1979,Reinders:1985,Narison:1989}. Contrary to the
important quark condensate $\langle 0 \mid : \bar{q}q : \mid 0
\rangle$ and gluon condensate $\langle 0 \mid
:G^{a}_{\mu\nu}G^{a}_{\mu\nu} : \mid 0 \rangle$, the quark gluon
mixed condensate $\langle 0
\mid:\bar{q}[ig_{s}G^{a}_{\mu\nu}\sigma_{\mu\nu}
\frac{\lambda^{a}}{2}] q : \mid 0 \rangle$ characterizes the direct
correlation between quarks and gluons, and together with the nonzero
two quark condensate $\langle 0 \mid :\bar{q}q :\mid 0 \rangle$, it
is responsible for the spontaneous breakdown of chiral symmetry. In
our previous works, we have studied the various non-perturbative
quantities at zero temperature by use of the DSEs in the " rainbow "
truncation, i.e. the quark condensate, the quark gluon mixed
condensate, susceptibility, and so on\cite{Zhou:2009ms}. Comparing
our theoretical results with others, such as, QCD sum
rules\cite{Kisslinger:2001}, Lattice QCD\cite{Takumi:2003}, we find
our calculations are in a good agreement with them. Now, we want to
extend the calculations of DSEs to nonzero temperature. As is known
to all, solve DSEs at finite temperature is quite difficult, but
with separable model interactions greatly simplifies the
calculations\cite{Burden:1996nh,Blaschke:2000gd}. In the present
work, we study the DSEs at finite temperature by use of the
separable model interactions. The main interesting of this work lies
on the consideration of nonzero temperature, which allow to study
the QCD phase diagram along the axis of zero chemical potential,
including deconfinement and the chiral symmetry restoration.

\section{Dyson-Schwinger Equations at zero and finite temperature}

\subsection{Dyson-Schwinger Equations at zero temperature}

To study quark and gluon condensates, we need to know the quark
propagators, which determine various quark condensates and the
quark gluon mixed condensates under the OPE constraints. The quark
propagator in configuration space is defined by
\begin{eqnarray}
S_{f}(x) = \langle0|T q(x)\bar{q}(0)|0\rangle .
\end{eqnarray}
For the physical vacuum, the quark propagator can be divided a
perturbative $S_{f}^{PT}(x)$ and a non-perturbative part
$S_{f}^{NPT}(x)$, one can write\cite{zhou:2004,Kisslinger:1998}
\begin{eqnarray}
S_{f}(x) = S_{f}^{PT}(x)+S_{f}^{NPT}(x).
\end{eqnarray}
In momentum space, $S_{f}^{NPT}(p)$ is related to the quark
self-energy, so the quark propagator of DSEs can be written
\begin{eqnarray}
S_{f}^{-1}(p) = i\gamma\cdot p+m_{f} + \frac{4}{3}g^{2}_{s}\int
\frac{d^{4}k}{(2\pi)^{4}} \gamma^{\mu} S_{f}(k) \Gamma^{\nu}(k, p )
G_{\mu\nu}(p-k) \label{equ:DSE}.
\end{eqnarray}
In Eq.(\ref{equ:DSE}), $g_{s}$ is the strong coupling constant of
QCD with the usual $\alpha_{s}(Q)$ by the relationship of $\alpha_{s}= g^{2}_{s}(Q)/4\pi$. The $G_{\mu\nu}(p-k)$ denotes fully dressed gluon propagator, and $m_{f}$ is the current quark mass with the subscript $f$ to stand for
quark flavor. In Feynman gauge, the simplest separable Ansatz has
following form\cite{Burden:1996nh,Blaschke:2000gd}
\begin{eqnarray}
g^{2}_{s}G_{\mu\nu}(p-k)\rightarrow \delta_{\mu\nu}
G(p^{2},k^{2},p\cdot k),
\end{eqnarray}
\begin{eqnarray}
G(p^{2},k^{2},p\cdot
k)=D_{0}F_{0}(p^{2})F_{0}(k^{2})+D_{1}F_{1}(p^{2})F_{1}(k^{2})(p\cdot
k),
\end{eqnarray}
where $D_{0}$ and $D_{1}$ are two strength parameters, and $F_{0}$
and $F_{1}$ are corresponding form factors.

As it is impossible to solve the complete set of DSE's, one has to
find a physically acceptable way to truncate this infinite tower and
make it soluble. To do it, we use a bare vertex $\gamma^{\nu}$ to
replace the full one $\Gamma^{\nu}(k, p)$ in Eq.(\ref{equ:DSE}).
This procedure is called as "Rainbow" approximation of DSEs. Thus,
Eq.(\ref{equ:DSE}) then becomes to
\begin{eqnarray}
S_{f}^{-1}(p) = i\gamma\cdot p+m_{f} + \frac{4}{3}g^{2}_{s}\int
\frac{d^{4}k}{(2\pi)^{4}} \gamma^{\mu} S_{f}(k) \gamma^{\nu}(k, p )
G_{\mu\nu}(p-k) \label{equ:DSE1}.
\end{eqnarray}

An important observation is that the general form of the inverse
quark propagator $S^{-1}_{f}(p)$ can be rewritten in Euclidean
space\cite{Robert:2000} as
\begin{eqnarray}
S^{-1}_ {f}(p) = i\gamma \cdot p  A_ {f}(p^{2}) + B_
{f}(p^{2})\label{equ:Sf},
\end{eqnarray}
with  $A_{f}$ and  $B_{f}$ are scalar functions of the $p^{2}$ .

With "Rainbow" truncation, we can obtain the coupling integral
equations for quark amplitudes $A_{f}(p^{2})$ and $B_{f}(p^{2})$ and these coupling equations take the form
in the Feynman gauge
\begin{eqnarray}
[A_{f}(p^{2})-1] p^{2} = \frac{8}{3} \int \frac{d^{4}q}{(2\pi)^{4}}
G(p-q)\frac{A_{f}(q^{2})}{q^{2}A^{2}_{f}(q^{2})+ B^{2}_{f}(q^{2})}p
\cdot q,
\end{eqnarray}
\begin{eqnarray}
B_{f}(p^{2})-m_{f}= \frac{16}{3} \int \frac{d^{4}q}{(2\pi)^{4}}
G(p-q) \frac{B_{f}(q^{2})}{q^{2}A^{2}_{f}(q^{2})+B^{2}_{f}(q^{2})}.
\end{eqnarray}

\subsection{Extension to Finite temperature}

So far, we only consider the quark propagator at zero temperature.
An extension to temperature dependence of the DSEs from the zero temperature to the finite
temperature is systematically accomplished by a transcription of the
Euclidean quark four momentum via $p \rightarrow p_{n}= (
\omega_{n},\vec{p})$, where $\omega_{n} = (2n+1) \pi T$ are the
discrete Matsubara frequencies\cite{Matsubara:1955}. Therefore, a
sum over the Matsubara frequencies replace the integral over the
energy.

The fully dressed quark propagator of DSEs at finite temperature can
be written
\begin{eqnarray}
S^{-1}_{f}( p_{n}, T) = i \vec{\gamma}\cdot \vec{p}
A_{f}(p^{2}_{n},T) + i \gamma_{4} \omega_{n} C_{f}(p^{2}_{n},T) +
B_{f}(p^{2}_{n},T),
\end{eqnarray}
where $p^{2}_{n} = \omega^{2}_{n} + \vec{p}^{2}$. Due to the
breaking of $O(4)$ symmetry in the four momentum space, we have
three quark amplitudes $A_{f}$, $B_{f}$ and $C_{f}$. The solutions
have the form $ A_{f}(p^{2}_{n},T) = 1 + a_{f}(T)F_{1}(p^{2}_{n})$,
$ B_{f}(p^{2}_{n},T) = m_{f} + b_{f}(T)F_{0}(p^{2}_{n})$ and
$C_{f}(p^{2}_{n},T)= 1 + c_{f}(T) F_{1}(p^{2}_{n})$, are defined by
the temperature dependent coefficients $a_{f}(T)$, $b_{f}(T)$ and
$c_{f}(T)$. The explicit form for $a_{f}(T)$, $b_{f}(T)$ and
$c_{f}(T)$ is given by
\begin{eqnarray}
a_{f}(T) = \frac{8D_{1}}{9} T \sum_{n} \int
\frac{d^{3}p}{(2\pi)^{3}}F_{1}(p^{2}_{n}) \vec{p}^{2}[ 1 + a_{f}(T)
F_{1}(p^{2}_{n})]d^{-1}_{f}(p^{2}_{n},T) \label{equ:af},
\end{eqnarray}

\begin{eqnarray}
c_{f}(T) = \frac{8D_{1}}{3} T \sum_{n} \int
\frac{d^{3}p}{(2\pi)^{3}}F_{1}(p^{2}_{n}) \omega^{2}_{n}[ 1 +
c_{f}(T) F_{1}(p^{2}_{n})]d^{-1}_{f}(p^{2}_{n},T) \label{equ:cf},
\end{eqnarray}

\begin{eqnarray}
b_{f}(T) = \frac{16D_{0}}{3} T \sum_{n} \int
\frac{d^{3}p}{(2\pi)^{3}}F_{0}(p^{2}_{n}) [ m_{f} + b_{f}(T)
F_{0}(p^{2}_{n})]d^{-1}_{f}(p^{2}_{n},T) \label{equ:bf},
\end{eqnarray}
where the denominator of the quark propagator $S_{f}(p_{n},T)$,
$d_{f}(p^{2}_{n},T)$, is given by
\begin{eqnarray}
d_{f}(p^{2}_{n},T) = \vec{p}^{2} A^{2}_{f}( p^{2}_{n}, T) +
\omega^{2}_{n}C^{2}_{f}( p^{2}_{n}, T) + B^{2}_{f}( p^{2}_{n}, T)
\label{equ:df}.
\end{eqnarray}
As we know, solve DSEs at finite temperature is quite difficult, but
with separable model interactions greatly simplifies the
calculations. For simplicity, we choose the following form for the
separable interaction form factor\cite{Horvatic:2007wu}:
\begin{eqnarray}
F_{0}(p^{2}) = exp ( -p^{2}/\Lambda^{2}_{0})\label{equ:f0},
\end{eqnarray}
\begin{eqnarray}
F_{1}(p^{2}) = \frac{1 + exp(-p^{2}_{0}/\Lambda^{2}_{1})}{1 +
exp((p^{2}-p^{2}_{0})/\Lambda^{2}_{1})}\label{equ:f1},
\end{eqnarray}
which is successful used to describe the phenomenology of the light
pseudoscalar mesons. Substituting Eqs.(\ref{equ:f0},\ref{equ:f1})
into Eqs.(\ref{equ:af}-\ref{equ:bf}) one can solving gap equations
for a given temperature $T$, and get the quark amplitudes $A_{f}$,
$B_{f}$ and $C_{f}$, but there is need to control the appropriate
number of Matsubara modes in calculation.

At the lowest dimension, quark and gluon condensates play essential
role in describing properties of nuclear matter and hadron
structure. The nonlocal quark condensate $\langle 0\mid
:\bar{q}(x)q(0):\mid 0\rangle$ can be given as
\cite{Kisslinger:1997qs}

\begin{eqnarray}
&&\langle 0\mid :\bar{q}(x)q(0):\mid 0\rangle~~~~~~~~~~~~~~~~~~~~~~~~~~~~~~~~~~~~~~~~~ \nonumber \\
&=&(-4N_{c})\int
\frac{d^{4}p}{(2\pi)^4}\frac{B_{f}(p^{2})e^{ipx}}{p^{2}A^{2}_
{f}(p^{2})+ B^{2}_{f}(p^{2})}~~~~~~~~~~~~ \nonumber\\
&=&-\frac{3}{4\pi^{2}}\int_{0}^{\infty}p^2 dp^2
\frac{B_{f}(p^2)}{p^2A^{2}_{f}(p^2) + B^{2}_{f}(p^2)}
\frac{2J_{1}(\sqrt{p^2x^2})}{\sqrt{p^2x^2}}\label{equ:nonlo1},
\end{eqnarray}
where $N_{c}=3$ is number of colors.  $J_{1}$ in  Eq.(\ref{equ:nonlo1}) is
Bessel function. When $x=0$ , the local quark vacuum condensate is
given by
\begin{eqnarray}
 \langle 0\mid : \bar{q}(0)q(0):\mid 0\rangle =
-4N_{c} \int \frac{d^{4}p}{(2\pi)^4} \frac{B_{f}(p^2)}{p^2 A^{2}_
{f}(p^2) + B^{2}_{f}(p^2)}\label{equ:qq}
\end{eqnarray}

Another important physical quantity is the four quark condensate.
The factorization hypothesis for the four quark condensate is
well-known from the works of M. A. Shifman, A. I. Vainshtein, and V.
I. Zakharov\cite{Shifman:1979}, and has been extensively used in QCD
sum rules through the operator product expansion approach.
 For the nonlocal four quark condensate$\langle 0\mid : \bar{q}(0)\gamma_{\mu}\frac{\lambda_C^a}{2}q(0)
 \bar{q}(0)\gamma_{\mu}\frac{\lambda_C^a}{2}q(0):\mid 0\rangle$, according to
 Ref.\cite{Meissner:1997,zong:1999,zong:2000}, we have
\begin{eqnarray}
 &&\langle 0\mid : \bar{q}(0)\gamma_{\mu}\frac{\lambda_C^a}{2}q(0)
 \bar{q}(0)\gamma_{\mu}\frac{\lambda_C^a}{2}q(0):\mid 0\rangle~~~~~~ \nonumber \\
 &=&-\int \frac{d^{4}p}{(2\pi)^4}\int\frac{d^{4}q}{(2\pi)^4}e^{ix.(p-q)}
[4^3\frac{B_{f}(p^2)}{p^2 A^{2}_{f}(p^2)+
B^{2}_{f}(p^2)}\frac{B_{f}(q^2)}{q^2 A^{2}_ {f}(q^2)+
B^{2}_{f}(q^2)}~~~~~ \nonumber \\
&+&32\frac{A_{f}(p^2)}{p^2 A^{2}_{f}(p^2)+
B^{2}_{f}(p^2)}\frac{A_{f}(q^2)}{q^2 A^{2}_{f}(q^2)+
B^{2}_{f}(q^2)}p\cdot q]\label{equ:nonqqqq}.
\end{eqnarray}

Similarly, the local ( $x=0$ ) four quark vacuum condensate is given
by
\begin{eqnarray}
 &&\langle 0\mid : \bar{q}(0)\gamma_{\mu}\frac{\lambda_C^a}{2}q(0)
 \bar{q}(0)\gamma_{\mu}\frac{\lambda_C^a}{2}q(0):\mid 0\rangle~~~~~~ \nonumber \\
 &=&-4^3 \int \frac{d^{4}p}{(2\pi)^4}
\left[\frac{B_{f}(p^2)}{p^2 A^{2}_{f}(p^2)+
B^{2}_{f}(p^2)}\right]^2~~~~~ \nonumber \\
&=&-\frac{4}{9} \langle 0\mid : \bar{q}(0)q(0):\mid 0\rangle^2
\label{equ:qqqq},
\end{eqnarray}
which is consistent with the vacuum saturation assumption of
Ref.\cite{Shifman:1979}.

Besides the quark condensate, the quark gluon mixed condensate is
another important chiral order parameter, which plays an important
role in QCD sum rules. In the frame work of the global color
symmetry model (GCM), the quark gluon mixed condensate are given
by\cite{Meissner:1997,Zhang:2004xg}
\begin{eqnarray}
 &&\langle 0\mid : \bar{q}(0)g\sigma\cdot G(0)q(0):\mid 0\rangle
 ~~~~~ \nonumber \\
 &=&-\frac{N_{c}}{16\pi^{2}}\{12\int dp^{2}
\frac{p^{4}B_{f}(p^2)(2-A_{f}(p^2))}{p^2 A^{2}_{f}(p^2)+
B^{2}_{f}(p^2)}~~~~~ \nonumber \\
&+&\frac{27}{4}\int dp^{2}
p^{2}B_{f}(p^2)\frac{[2A_{f}(p^2)(A_{f}(p^2)-1)]p^2+B^2_{f}(p^2)}{p^2
A^{2}_{f}(p^2)+ B^{2}_{f}(p^2)} \}\label{equ:qgq}
\end{eqnarray}

It is common belief that the quark condensate, which determines
light quark mass, depends on temperature T. In the case of finite
temperature, one usually takes the same expression to study the
temperature dependence of the quark
condensate\cite{Blaschke:2000gd}, we have then
\begin{eqnarray}
 \langle 0\mid : \bar{q}(0)q(0):\mid 0\rangle _{T}=
-4N_{c}T\sum_{n=-\infty}^{\infty}\int \frac{d^{3}p}{(2\pi)^3}
\frac{B_{f}(p^{2}_{n}, T)}{\vec{p}^2A^{2}_{f}(p^{2}_{n},T) +
\omega^2_{n}C^{2}_
{f}(p^{2}_{n},T)+B^{2}_{f}(p^{2}_{n},T)}\label{equ:qq-T}
\end{eqnarray}

 It is still a matter of debate for the four quark condensate when
 $T\neq0$. It was shown in Ref.\cite{Narison:1983}, that
 factorization hypothesis implies that the four quark condensate becomes
 dependent on the QCD renormalization scale. In addition,
 theoretical arguments from the chiral perturbation theory also
 do not support this approximation at next to next to leading order,
 except in the chiral limit\cite{GomezNicola:2010tb,GomezNicola:2012uc}.
For simplicity, we takes the form
\begin{eqnarray}
 &&\langle 0\mid : \bar{q}(0)\gamma_{\mu}\frac{\lambda_C^a}{2}q(0)
 \bar{q}(0)\gamma_{\mu}\frac{\lambda_C^a}{2}q(0):\mid 0\rangle_{T}
=-\frac{4}{9} \langle 0\mid : \bar{q}(0)q(0):\mid 0\rangle_{T}^2
\label{equ:qqqq-T},
\end{eqnarray}

As to the quark gluon mixed condensate in $T\neq 0$ region, we have
\begin{eqnarray}
 &&\langle 0\mid : \bar{q}(0)g\sigma\cdot G(0)q(0):\mid 0\rangle
 _{T}~~~~~ \nonumber \\
 &=&-36T\sum_{n=-\infty}^{\infty}\int \frac{d^{3}p}{(2\pi)^3}\{
\frac{B_{f}(p^2)[(2-A_{f}(p^2))p^2+(2-C_{f}(p^2))\omega^2_{n}]}{p^2
A^{2}_{f}(p^2)+C_{f}(p^2)\omega^2_{n}+
B^{2}_{f}(p^2)}~~~~~ \nonumber \\
&-&\frac{81}{4}T\sum_{n=-\infty}^{\infty}\int
\frac{d^{3}p}{(2\pi)^3}
\frac{2B_{f}[A_{f}(p^2)(A_{f}(p^2)-1)p^2+C_{f}(p^2)(C_{f}(p^2)-1)\omega^2_{n}]+B^3_{f}(p^2)}{p^2
A^{2}_{f}(p^2)+C_{f}(p^2)\omega^2_{n}+B^{2}_{f}(p^2)}
\}\label{equ:qgq}
\end{eqnarray}

The DSEs provide a valuable non-perturbation, renormalisable,
continuum tool for studying temperature dependent field theories. A
number of physical phenomena such as confinement, dynamical chiral
symmetry breaking, and temperature dependent of quark mass, which
cannot be explained by perturbation treatments, can be understood in
terms of its solution of the DSEs at zero and finite temperature.
Eq.(\ref{equ:af}-\ref{equ:bf}).

\section{Calculations and results }

\begin{figure}
\begin{center}
\includegraphics[width=12cm]{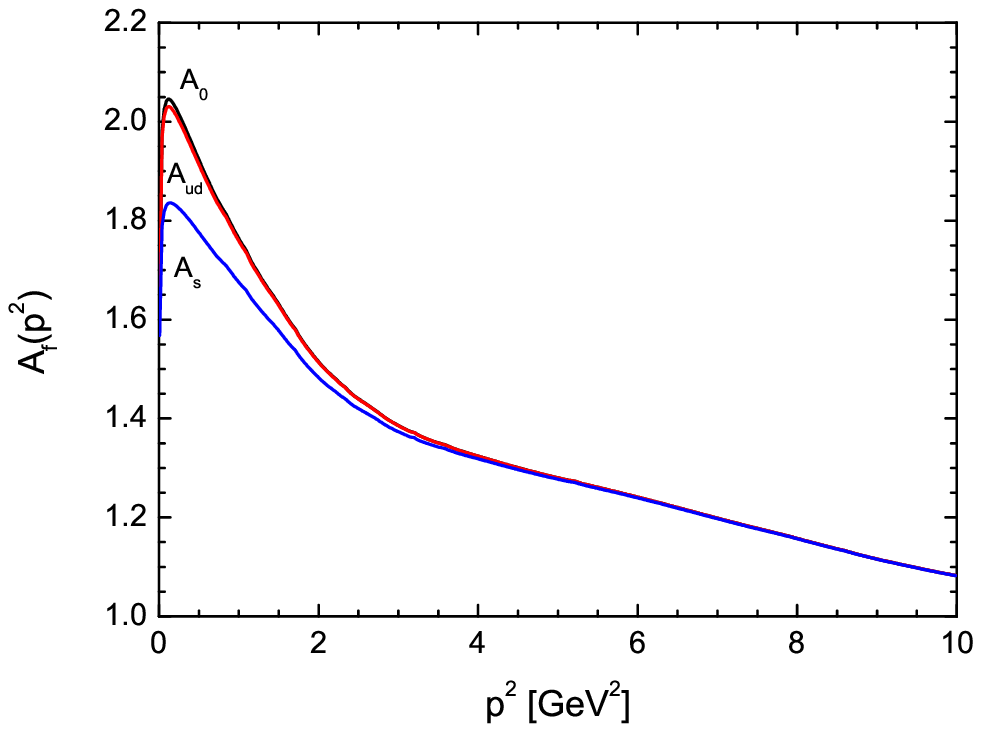}
\caption{$p^{2}$-dependence of quark self-energy amplitudes
$A_{f}(p^{2})$, subscript $f$ for the ud quark, the s quark and the
chiral limit cases.} \label{fig:A-p}
\end{center}
\end{figure}

\begin{figure}
\begin{center}
\includegraphics[width=12cm]{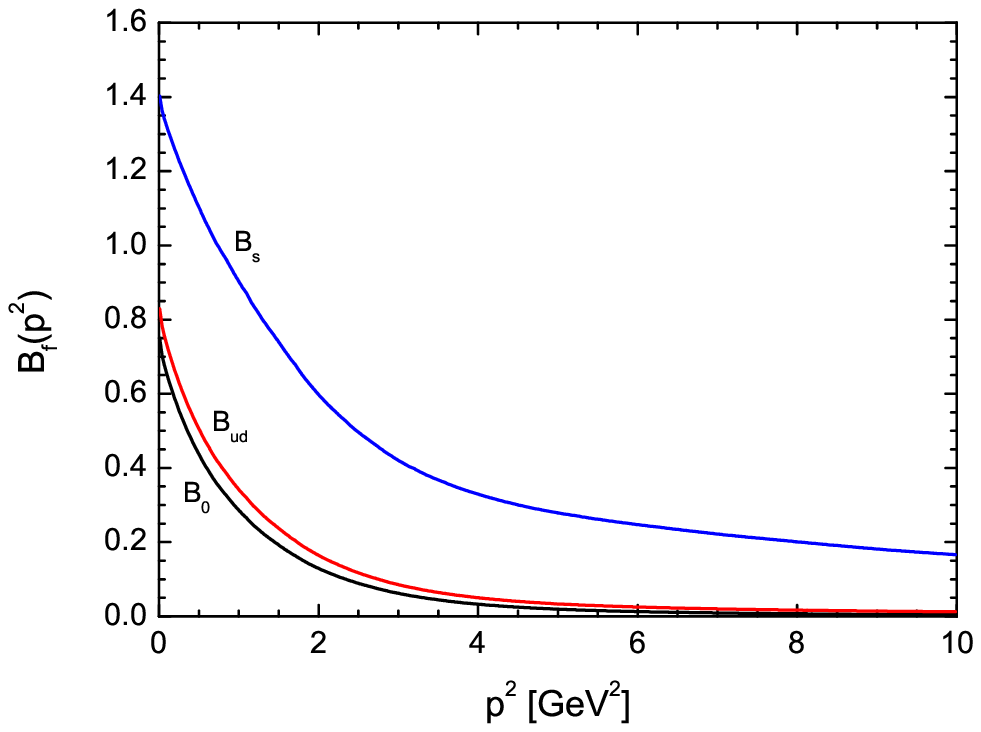}
\caption{$p^{2}$-dependence of quark self-energy amplitudes
$B_{f}(p^{2})$, subscript $f$ for the ud quark, the s quark and the
chiral limit cases.} \label{fig:B-p}
\end{center}
\end{figure}

We first solve the quark's DSEs at zero temperature with parameters
$m_{ud}=5.5MeV$, $m_{s}=117MeV$, $\Lambda_{0} = 758 MeV$,
$\Lambda_{1} = 961 MeV$, $p_{0} = 600 MeV$, $D_{0}\Lambda^{2}_{0} =
219$, $D_{1}\Lambda^{4}_{0} = 40$, which are completely fixed by
meson phenomenology calculated from the model as given in
\cite{Blaschke:2000gd}. The obtained results of DSEs at zero
temperature are displayed in Fig.\ref{fig:A-p} and
Fig.\ref{fig:B-p}. At the same time, we obtain the quark condensate
$\langle\bar{q}q\rangle=(0.202GeV)^{3}$ and the quark gluon mixed
condensate $\langle\bar{q}[g\sigma G ]q\rangle=(0.455GeV)^{5}$ at
T=0.

\begin{figure}
\begin{center}
\includegraphics[width=12cm]{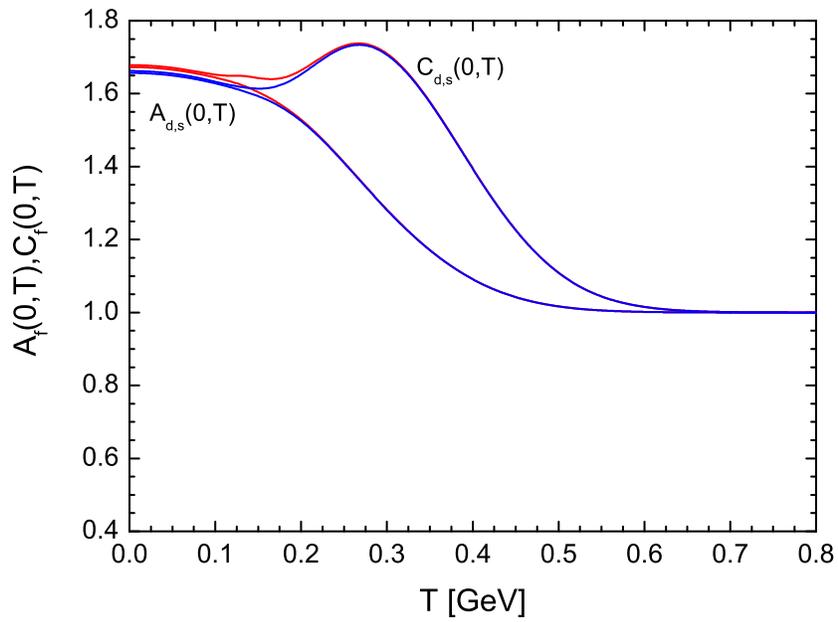}
\caption{T-dependence of quark self-energy amplitudes $A_{f}(0,T)$,
 and $C_{f}(0,T)$, subscript $f$ for the ud quark, the s
quark and the chiral limit cases. }\label{fig:AC-T}
\end{center}
\end{figure}

\begin{figure}
\begin{center}
\includegraphics[width=12cm]{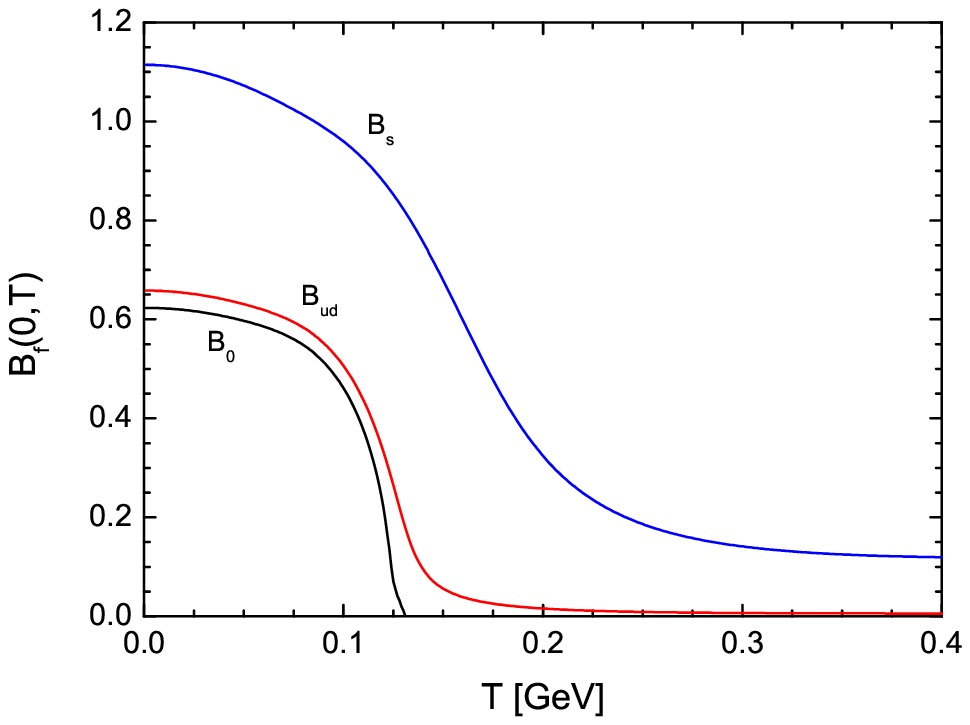}
\caption{T-dependence of quark self-energy amplitudes $B_{f}(0,T)$,
 subscript $f$ for the ud quark, the s quark and the chiral limit
cases. } \label{fig:B-T}
\end{center}
\end{figure}

In order to demonstrate the temperature dependence of quark
propagators, we use Matsubara formula, and we then solve quark's
DSEs at nonzero temperature with the same gluon propagator and
parameters. The results are given in Fig.\ref{fig:AC-T} and
Fig.\ref{fig:B-T}. From Fig.\ref{fig:AC-T}, we can find that, for
low temperature, the vector parts of the quark propagator
$A_{f}(0,T)$ and $C_{f}(0,T)$ coincide with each other, they are
almost the same. However, for the temperature higher than about
$T=131MeV$, they become distinctly different. That means the $O(4)$
symmetry has be broke.

\begin{figure}
\begin{center}
\includegraphics[width=12cm]{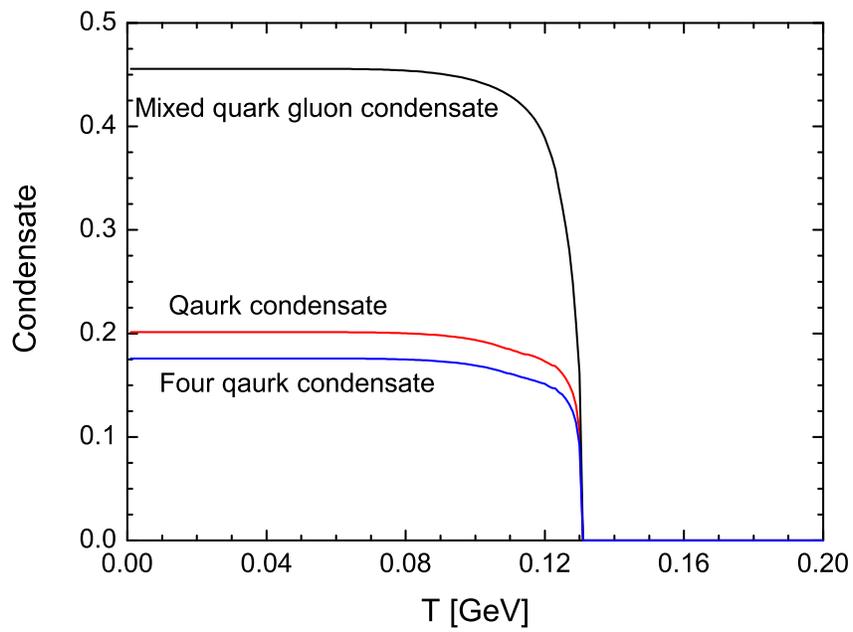}
\caption{T-dependence of the quark condensate, the four quark
condensate and the quark gluon mixed condensate in the chiral limit
cases.} \label{fig:locond-T}
\end{center}
\end{figure}

\begin{figure}
\begin{center}
\includegraphics[width=12cm]{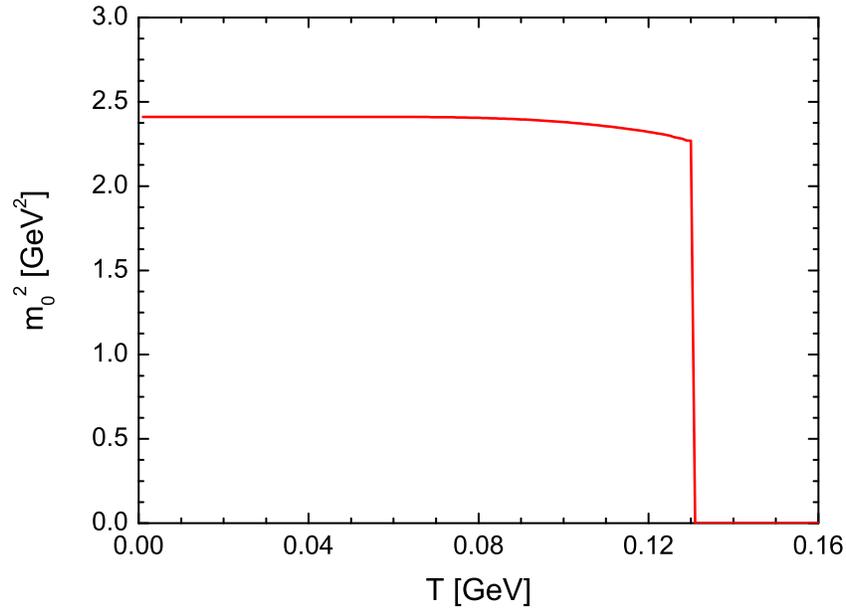}
\caption{T-dependence of the ratio of the quark gluon mixed
condensate to the quark condensate, $m_{0}^{2}=\frac{\langle 0\mid :
\bar{q}(0)g\sigma\cdot G(0)q(0):\mid 0\rangle
 _{T}}{\langle\bar{q}q\rangle_{T}}$,
in the chiral limit cases.}\label{fig:ratio-T}
\end{center}
\end{figure}

Using the individual solutions of the quark's DSEs at zero- and
finite temperature, $A_{f}$, $B_{f}$ and $C_{f}$ we respectively
obtain the properties of the QCD vaccum at zero- and nonzero
temperature in the chiral limit case. The quark condensate
$\langle\bar{q}q\rangle$, the four quark condensate
$\langle\bar{q}\Gamma q \bar{q}\Gamma q\rangle$ and the quark gluon
mixed condensate $\langle\bar{q}[g\sigma G ]q\rangle$ are important
condensates of the lowest dimension, which reflect the
non-perturbative structure of QCD vacuum state, and can be the
chiral order parameter of QCD. In Fig.\ref{fig:locond-T}, the
temperature dependence of the two quark-, the four quark- condensate
and the quark gluon mixed condensate in chiral limit in the
separable model are plotted respectively. The critical temperature
for the chiral symmetry restoration comes out to be $T_{c}=131MeV$.
These order parameters give a same critical temperature and the same
critical behavior. From Fig.\ref{fig:locond-T}, we find these three
condensates are almost independent of the temperature below $T_{c}$,
while a clear signal of chiral symmetry restoration is shown up at
$T_{c}$. We also calculate the ratio of the quark gluon mixed
condensate to the two quark condensate. The results are shown in
Fig.\ref{fig:ratio-T}. From the figure, we can see, although there
are drastic changes of the quark condensate and the quark gluon
mixed condensate near $T_{c}$, the ratio $m_{0}^{2}(T)$ is almost
flat when temperature at the region from 0 to $T_{c}$. For vacuum
state at the chiral limit, the ratio $m_{0}^{2}(0)=2.41GeV^{2}$,
which is larger in comparison with the results from lattice QCD
which is about $1 GeV^{2}$\cite{Doi:2004jp}, and suggests the great
significance that the quark gluon mixed condensate plays in OPE
calculations.

In summary, we study fully dressed quark propagator $S_{f}(p^{2},T)$
in QCD by using of the DSEs with zero and finite temperature under
the " rainbow " truncation, $\Gamma^{\nu}=\gamma^{\nu}$. We solve
the DSEs numerically and get quark propagator functions,
$A_{f}(p^{2},T)$, $B_{f}(p^{2},T)$ and $C_{f}(p^{2},T)$ in Eq.(10)
at two cases of $T=0$ and $p^{2}=0$, and then we obtained the quark
propagator $S_{f}(p^2,T)$. The resulting quark propagator has no
Lehmann representation and hence there are no quark production
thresholds in any calculations of observable. The absence of such
thresholds admits the interpretation that $S_{f}(p^{2},T)$ describes
the propagator of a confined quark. With the solutions of the
quark's DSEs $A_f$, $B_f$ and $C_f$, the temperature dependence of
the two quark condensate, the four quark condensate and the quark
gluon mixed condensate in the chiral limit are obtained. We find
these condensates have same critical temperature for the chiral
symmetry restoration and same critical behavior for QCD phase
transition, though which characterize different aspects of QCD
vacuum. At the same time, we study the ratio between quark gluon
mixed condensate and the two quark condensate, and obtain the
nontrivial result that the ratio is insensitive to temperature below
the critical point $T_c$.

\newpage

\newpage

\end{document}